\documentclass[aps,prb,showkeys,floatfix,amsmath,superscriptaddress, amssymb,reprint]{revtex4-1}
\usepackage{graphicx} 
\usepackage{bm}% bold math    
\usepackage{color,mathtools,bbm} 
\usepackage{wasysym}    
  \usepackage{lineno}
\usepackage{longtable}
\usepackage{footnotehyper}
\makesavenoteenv{longtable}
\usepackage{xcolor}
\usepackage{soul}
\usepackage{array}
\usepackage[normalem]{ulem}
\usepackage{hypernat}
\usepackage{nameref}
\usepackage{wrapfig} 
\usepackage{adjustbox}
\usepackage{multirow}
\usepackage{braket}
\usepackage{hyperref}
\begin{document}

\title{Tunable topological phase in 2D ScV$_6$Sn$_6$ kagome material}

%Authors
\author{Chidiebere I. Nwaogbo}
\altaffiliation{cin221@lehigh.edu}
\affiliation{Department of Physics, Lehigh University, Bethlehem, PA 18015, USA}
\author{Sanjib K. Das}
\affiliation{Department of Physics, Lehigh University, Bethlehem, PA 18015, USA}
\affiliation{Department of Physics and Astronomy, University of Delaware, Newark, DE 19716, USA}
\author{Chinedu E. Ekuma}
\affiliation{Department of Physics, Lehigh University, Bethlehem, PA 18015, USA}

\begin{abstract}
We investigate the topological properties of the vanadium-based 2D kagome metal ScV$_6$Sn$_6$, a ferromagnetic material with a magnetic moment of 0.86 $\mu_B$ per atom. Using \textit{ab initio} methods, we explore spin-orbit coupling-induced gapped states and identify multiple Weyl-like crossings around the Fermi energy, confirming a Chern number $|C| = 1$ and a large anomalous Hall effect (AHE) of 257 $\Omega^{-1}$cm$^{-1}$. Our calculations reveal a transition from a topological semimetal to a trivial metallic phase at an electric field strength of $\approx$0.40 eV/\text{\AA}. These findings position 2D ScV$_6$Sn$_6$  as a promising candidate for applications in modern electronic devices, with its tunable topological phases offering the potential for future innovations in quantum computing and material design.
\end{abstract}

\maketitle

\section{\label{sec:level1}INTRODUCTION}

The study of exotic phenomena in kagome materials has progressed significantly with the development of topological quantum matter theories \cite{hasan2015topological}. These advances have facilitated the discovery of various novel topological materials, including topological insulators (TIs) and topological semimetals (TSMs). Both classes exhibit unique features, such as insulating bulk states and boundary states protected by symmetries \cite{RevModPhys.82.3045,kane2005quantum,niu2015two,wang2013organic}. These boundary states often consist of Dirac or Weyl points, which connect energy bands of opposite parity. In the presence of spin-orbit coupling (SOC), the degeneracy at these points can be lifted, forming a topological gap. Notably, these intrinsic properties do not require external fields to manifest.

Nonmagnetic TIs and TSMs are characterized by topological states protected by time-reversal symmetry (TRS), while magnetic TSMs exhibit unique phases, such as Weyl nodes and Chern numbers, arising from the breaking of crystal symmetries \cite{okada2013observation}. Despite significant advances in understanding nonmagnetic topological phases, both insulating and (semi)metallic—magnetic topological materials remain  relatively less explored. The complex interactions within these materials pose a greater challenge for theoretical predictions compared to their nonmagnetic counterparts. However, they hold considerable technological potential due to the additional degrees of freedom provided by magnetism, which enable novel ways of manipulating topological states. Recent theoretical and experimental advances have deepened our understanding of magnetic topological materials \cite{xu2020high,watanabe2018structure,otrokov2019prediction}.

A key topological invariant in such systems is the Chern number (\textit{C}), which breaks TRS while preserving particle-hole symmetry. The Chern number is given by the integral of the Berry curvature over the Brillouin zone and corresponds to the number of stable chiral modes in the occupied bulk states \cite{sheng2006quantum}. In two-dimensional (2D) materials, a non-trivial value of \textit{C} indicates a 2D Chern insulator, characterized by delocalized boundary states at the edges of the material \cite{prodan2010entanglement,ahsan2023prediction}. Another important topological invariant is the $\mathbb{Z}_2$ invariant, denoted as $(v_0, v_1, v_2, v_3)$, used to classify both 2D and three-dimensional (3D) topological insulators \cite{hasan2015topological}. The $\mathbb{Z}_2$ invariant distinguishes trivial insulators ($\mathbb{Z}_2 = 0$) from topological insulator states ($\mathbb{Z}_2 = 1$), with a nontrivial $\mathbb{Z}_2$ associated with the quantum spin Hall (QSH) phase. The topological phase is classified as weak ($v_0 = 0$) or strong ($v_0 = 1$), depending on the value of $v_0$ \cite{ahsan2023prediction}.

Recent studies have predicted strong nonmagnetic topological insulator phases in kagome materials, which are known for their complex electronic properties, including symmetry-breaking charge density waves (CDWs) and intricate phase diagrams \cite{teng2022discovery,tan2021charge,yu2021concurrence,hu2022tunable}. For instance, materials such as GdV$_6$Sn$_6$ and YV$_6$Sn$_6$ demonstrate behavior akin to topological insulators in their paramagnetic states \cite{pokharel2021electronic,hu2022tunable}. Similarly, bulk TbV$_6$Ge$_6$, a member of the RV$_6$Ge$_6$ family, has been predicted to exhibit topological crystalline insulator properties, with these topological features protected by its structural symmetry \cite{ahsan2023prediction}. In the realm of 2D materials, the study of magnetic topological insulators and semimetals is just beginning, with only a few examples, such as MnTe(Bi$_2$Te$_3$) and MnBi$_2$Te$_4$, identified to date. MnBi$_2$Te$_4$ exhibits antiferromagnetic behavior, with a Néel temperature (\textit{T}$_N$) of approximately 25 K \cite{otrokov2019prediction}. Other notable discoveries include the kagome ferromagnetic (FM) Weyl semimetal Co$_3$Sn$_2$S$_2$ and the magnetically induced Weyl semimetal GdPtBi \cite{morali2019fermi,hirschberger2016chiral}.

A series of magnetic topological semimetals with significant experimental potential have been predicted and identified, driven by searches for large anomalous Hall effects (AHE) and \textit{ab initio} calculations \cite{kubler2014non, zhang2017strong}. The anomalous Hall effect, first discovered by Edwin Hall, occurs in all ferromagnetic semimetals and metals, where the Hall resistivity's response to an applied magnetic field mirrors the magnetization response \cite{aliev2019novel}. The AHE is now understood to be significantly influenced by the Berry curvature, which acts like a magnetic monopole at energy level crossings, also known as Weyl points \cite{Berry1984quantal}. While most studies on TIs and TSMs have focused on intrinsic topological properties, recent research has explored the tunability of these phases through external perturbations. For example, applying external electric fields (\textit{E}-fields) can shift the position of Dirac cones in momentum space, inducing topological phase transitions. In bilayer Bi(111), an infinitesimal \textit{E}-field can transition the material from a non-trivial to a trivial state \cite{Sawahata_2019}. Similarly, phosphorene, a 2D material derived from black phosphorus, can transition from a normal insulator to a topological insulator and eventually to a metal under an applied electric field \cite{liu2015switching}. Other studies have also identified tunable topological phases in 2D magnetic materials. For example, monolayer MoO which exhibits anomalous Hall states under shear strain, and  single-layer RuClBr which hosts a quantum anomalous valley Hall phase with chiral spin-valley locking \cite{sun2022valley, wu2023quantum}. These findings underscore the potential of 2D materials to host controllable topological phases through external fields.

Motivated by the emerging topological properties that characterize low-dimensional systems, we investigated the electronic structure and topological features of the two-dimensional kagome material ScV$_6$Sn$_6$, a member of the HfFe$_6$Ge$_6$ family. In its bulk form, ScV$_6$Sn$_6$ is a nontrivial metal exhibiting a tunable charge density wave (CDW) state with a first-order phase transition and a propagation vector of $\left(\tfrac{1}{3}, \tfrac{1}{3}, \tfrac{1}{3}\right)$, as determined from x-ray and neutron scattering experiments~\cite{arachchige2022charge,destefano2023pseudogap}. In this work, we explore the emergent properties of its 2D counterpart. Our \textit{ab initio} simulations indicate that 2D ScV$_6$Sn$_6$ is a topological material with a ferromagnetic ground state characterized by a mean-field Curie temperature of $T_C \approx 89\,$K and a magnetic moment of approximately $0.86\,\mu_B$ per atom. This behavior contrasts with the nonmagnetic nature of its three-dimensional bulk form. The material is further characterized by Weyl points formed by pairs of topological bands near the Fermi energy, with a Chern number of $|C| = 1$. Furthermore, a significant anomalous Hall conductivity is observed at the Fermi level, related to the ferromagnetic ordering in the system~\cite{chang2023colloquium,vzelezny2023high,ye2018massive}. These chiral states, which break time-reversal symmetry, classify 2D ScV$_6$Sn$_6$ as a magnetic Weyl semimetal with nontrivial topology. Additionally, we demonstrate that the nontrivial topological phase remains robust under the application of moderate electric fields, but transitions from a nontrivial to a trivial phase at an electric field strength of $\approx 0.4\,$eV/\AA.

\begin{figure}
    
    \centering
    \begin{minipage}{.5\textwidth}
        \includegraphics[width=\textwidth]{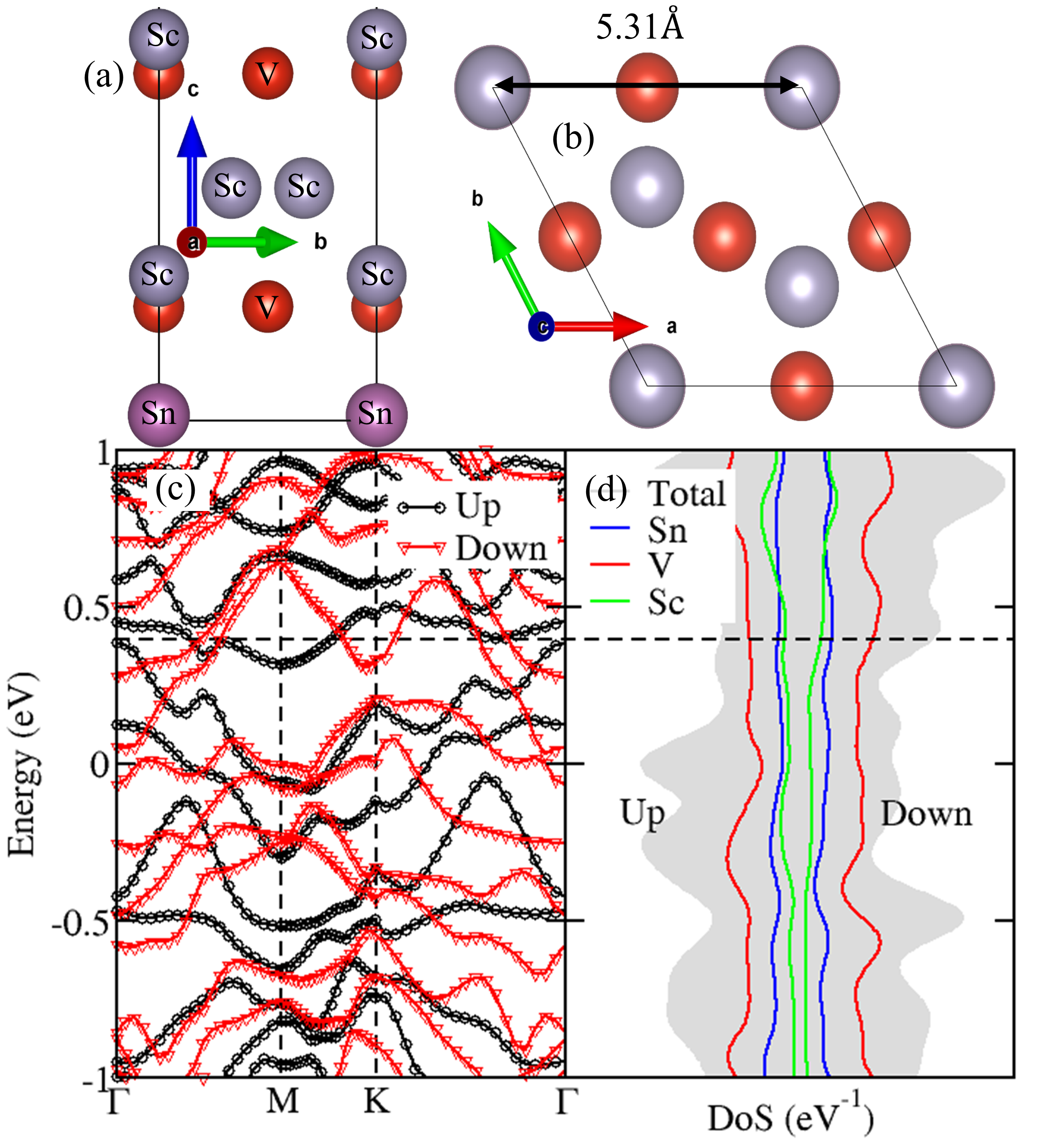}
       \caption{(a) Side and (b) top views of the crystal structure of 2D ScV$_6$Sn$_6$. (c) Electronic band structure with the spin-up states (black bubbles) and spin-down states (red triangles). (d) Projected density of states onto atomic orbitals. The blue, red, and green lines represent the contributions from Sn, V, and Sc atoms, respectively, while the gray shading denotes the total density of states for the 2D structure. The left panel of (d) shows spin-up states, and the right panel shows spin-down states. The horizontal dotted line indicates the Fermi energy at $\approx0.40\,$ eV.}
        \label{fig:structureband}
    \end{minipage}
    
\end{figure}

\section{\label{sec:level2}COMPUTATIONAL METHODS}

The bulk structure of ScV$_6$Sn$_6$ crystallizes in the hexagonal space group $P6/mmm$ (No. 183) and comprises two layers of vanadium (V) kagome nets, each formed by six V atoms. To investigate its two-dimensional (2D) properties, a monolayer was cleaved from the bulk structure, and a vacuum layer of approximately 23 \AA~ was introduced along the c-axis to eliminate interactions between periodic images. The resulting 2D structure preserves the hexagonal symmetry of the bulk, with lattice constants $a = b = 5.31$ \AA~ (see Figure \ref{fig:structureband}(a)-(b)). Density functional theory (DFT)~\cite{kohn1996density} calculations were performed using the Vienna Ab initio Simulation Package (VASP) \cite{kresse1996efficiency,kresse1996efficient}, employing the generalized gradient approximation (GGA) with the Perdew-Burke-Ernzerhof (PBE) exchange-correlation functional~\cite{perdew1996generalized}. Structural relaxation was carried out until the total energy and atomic forces converged to $\approx 10^{-7}$ eV and 0.02 eV/\AA, respectively. A $\Gamma$-centered $5 \times 5 \times 1$ k-point mesh and a plane-wave energy cutoff of 500 eV were used for these calculations. For the electronic structure analysis, we employed first-principles, spin-polarized DFT calculations with and without an effective on-site Coulomb interaction (DFT+\textit{U}), including spin–orbit coupling (SOC) to capture relativistic effects. To assess the role of electronic correlations, we varied the Hubbard \textit{U} parameter from 0 to 6.0~eV for both V and Sc atoms, with the upper bound chosen in accordance with previous studies on related systems~\cite{anisimov2005full,BERDECIA2024113067}. To explore the topological properties of 2D ScV$_6$Sn$_6$, maximally localized Wannier functions (MLWFs) \cite{Marzari_2012, prodan2010entanglement} were employed to construct a low-energy effective Hamiltonian using the Wannier90 code~\cite{Pizzi2020}. The Wannierization process included Sc-$d$, V-$d$, and Sn-$p$ orbitals, as they contribute significantly to the electronic states near the Fermi level. Topological characteristics, including Chern numbers and edge states, were computed using the WannierTools package~\cite{WU2017}. A semi-infinite system was modeled with a surface slab of $\approx$250 \AA~ thick, to ensure that the top and bottom surfaces are sufficiently decoupled, preventing artificial hybridization between edge states on opposite sides. Then the surface Green’s function was calculated to derive the local density of states and edge state dispersion.  The robustness of the topological phase was further examined by applying an external electric field. The intrinsic anomalous Hall conductivity (AHC) was computed by integrating the Berry curvature over the entire Brillouin zone for all occupied bands, offering insight into the system’s topological response under varying conditions.

\section{\label{sec:level3}RESULTS AND DISCUSSION}
We begin our analysis by establishing the stability of 2D ScV$_6$Sn$_6$, evaluating its formation energy ($E_{form}$), cleavage energy ($E_{cl}$), and mechanical and dynamical stability. The formation energy, calculated using $E_{form} = E_{tot} - \sum n \epsilon_i$, where $E_{tot}$ is the ground-state energy of the optimized 2D ScV$_6$Sn$_6$ and $\epsilon_i$ are the ground-state energies of the constituent elements in their bulk states, is $E_{form} \approx -0.17~\text{eV}$. This exothermic value indicates that the 2D phase is stable relative to the elemental state. The cleavage energy, calculated as $E_{cl} = (E_{l} - E_{bulk})/A$, where $E_{l}$ and $E_{bulk}$ are the energies of the cleaved layer and bulk structure, respectively, and $A$ is the surface area, was found to be $E_{cl} \approx 0.27~\text{J/m}^2$. This value is comparable to those of other experimentally synthesized 2D materials, such as MoS$_2$ ($0.09 - 0.22 ~\text{J/m}^2$), graphene ($0.19 - 0.72 ~\text{J/m}^2$), and significantly lower than that of black phosphorus ($\approx 0.37 ~\text{J/m}^2$), suggesting weak interlayer interactions and high experimental synthesizability~\cite{wang2015measurement, coleman2011two,tang2014nanomechanical,gaur2014surface}.

To evaluate the mechanical stability and probe the low-energy phonon responses, we performed 2D planar elastic constant calculations using the ElasTool toolkit~\cite{liu2022elastool,EKUMA2024109161}. The computed elastic constants satisfy the Born stability conditions specific to hexagonal symmetry~\cite{born1996dynamical,PhysRevB.90.224104}, thereby confirming the material’s mechanical stability. The elastic tensor exhibits eigenvalues of $36.13~\text{N/m}$, $72.26~\text{N/m}$, and $227.26~\text{N/m}$, from which we extract a 2D Young’s modulus of $109.66~\text{N/m}$ and a shear modulus of $36.13~\text{N/m}$. These quantities reflect the material’s substantial resistance to both axial and shear deformations. Furthermore, a Poisson's ratio of $0.52$ and a Pugh’s modulus ratio of $3.14$ place ScV$_6$Sn$_6$ firmly in the ductile regime, a classification corroborated by the computed strain energy density of $1.023~\text{J/m}^2$. To probe dynamical stability, we analyzed the material’s acoustic phonon responses, which are intrinsically linked to the elastic behavior under long-wavelength, low-frequency perturbations. The sound velocities—$V_l = 4.59~\text{km/s}$ (longitudinal), $V_s = 2.26~\text{km/s}$ (shear), and the average $V_d = 2.86~\text{km/s}$—directly correspond to the slopes of the acoustic phonon branches near the $\Gamma$ point. These modes govern the stress-strain response in the low-energy limit and are thus a microscopic reflection of the macroscopic elastic constants. The calculated Debye temperature of $T_D = 560.61~\text{K}$ further supports the material's vibrational stability and low-energy phonon rigidity. Taken together, the favorable elastic constants, high sound velocities, and elevated Debye temperature provide compelling evidence for the dynamical stability of 2D ScV$_6$Sn$_6$—a critical attribute for its potential integration into flexible and robust quantum materials platforms.

We investigate the electronic properties of 2D ScV$_6$Sn$_6$ by performing spin-polarized electronic structure calculations, including band structure and density of states analyses, as shown in Figure~\ref{fig:structureband}(c)–(d). The results demonstrate that 2D ScV$_6$Sn$_6$ is metallic, exhibiting multiple Weyl-like band crossings near the Fermi energy. All topological calculations presented were performed without the inclusion of a Hubbard $U$ correction. To verify the robustness of the observed electronic and topological features, we subsequently varied the Hubbard $U$ parameter applied to the V 3\textit{d} and Sc orbitals from 0 to 6~eV (see Supplementary Figure~\cite{supp}). This analysis confirms that while a finite $U$ stabilizes a ferromagnetic ground state, the essential features of the electronic structure, particularly the metallic character and the Weyl-like crossings, remain qualitatively unchanged across this range. This robustness underscores the intrinsic nature of the topological metallic state in 2D ScV$_6$Sn$_6$.

Notably, the electronic states near the Fermi level are primarily derived from the V 3\textit{d} orbitals, with secondary yet non-negligible contributions from the Sn and Sc atoms. This orbital character is consistent with prior studies on the bulk counterpart of ScV$_6$Sn$_6$~\cite{ahsan2023prediction, di2023electronic}. Our spin-polarized calculations further reveal that the ferromagnetic configuration is the ground state, with an estimated Curie temperature of \(T_c \approx 89~\text{K}\), obtained via mean-field approximation. This result implies that the monolayer system undergoes spontaneous magnetic ordering and breaks time-reversal symmetry below \(T_c\), in contrast to the nonmagnetic behavior reported for the bulk compound~\cite{yi2024quantum, yi2024tuning}. To further elucidate the nature of magnetic anisotropy, we computed the magnetocrystalline anisotropy energy (MAE), defined as $\mathrm{MAE} = E_{\parallel} - E_{\perp}$, where \(E_{\parallel}\) and \(E_{\perp}\) are the total energies corresponding to in-plane and out-of-plane magnetization directions, respectively. The calculated MAE of 2.01~meV per formula unit is positive, indicating a lower energy for out-of-plane spin alignment. This establishes that the magnetic easy axis is oriented perpendicular to the plane, a feature that could play a critical role in stabilizing topologically nontrivial spin textures in the 2D system.

\begin{figure}[b!]
    \centering
    \begin{minipage}{.50\textwidth}
        \includegraphics[width=\textwidth]{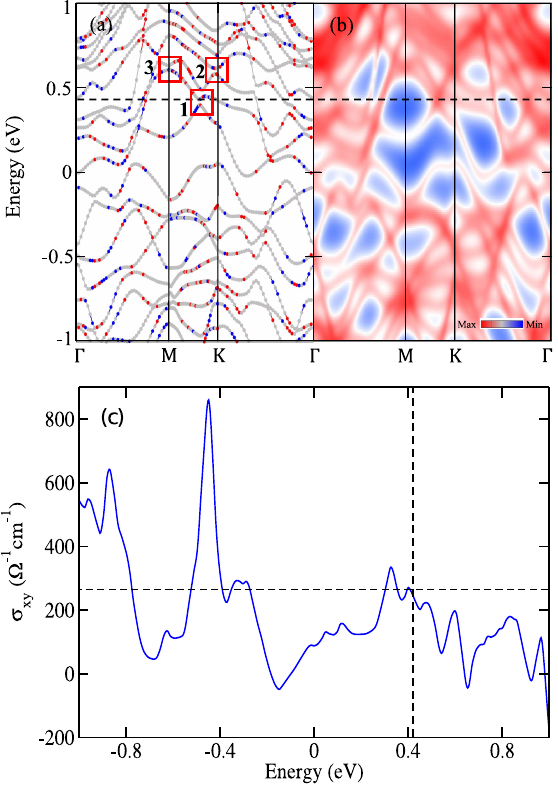}
      \caption{(a) Spin–orbit-coupling (SOC)–resolved electronic band structure of the 2D ScV$_6$Sn$_6$ slab, showing momentum-resolved orbital contributions. (b) Edge state dispersion along the [110] edge, with red regions indicating higher spectral weight and blue indicating lower weight. (c) Intrinsic anomalous Hall conductivity (AHC), $\sigma_{xy}$ (in $\Omega^{-1}~\text{cm}^{-1}$), as a function of energy. The horizontal dashed lines in (a) and (b), and the vertical dashed line in (c), indicate the Fermi energy at $\approx 0.42\,$eV. The horizontal dashed line in (c) marks the corresponding value of $\sigma_{xy} \sim 257\,\Omega^{-1}~\text{cm}^{-1}$ at the Fermi energy.}
        \label{fig:surface&AHC}
         \end{minipage}
\end{figure}

To explore the topological features of 2D ScV$_6$Sn$_6$, we analyze the edge states in the vicinity of the Fermi energy, which is set at \(E_f = 0.42\,\text{eV}\). These edge states originate from three gapped Weyl points (labeled 1–3 in Figure~\ref{fig:surface&AHC}(a)) as well as a spin–orbit coupling (SOC)-induced band gap centered at the Fermi level. These gaps belong to bands 55, 56, and 57, with energies of 33.0, 32.0, and 24.0 meV, respectively, with respect to the Fermi energy. We note that a Weyl point is only useful if it is located near the Fermi level, as only states at or near the Fermi energy can exhibit observable effects. Otherwise, contributions from the Weyl nodes may be obscured by other states. In the case of 2D ScV$_6$Sn$_6$, all the three gaps are positioned near the Fermi energy. Notably, Weyl gap 1 is exactly at the Fermi energy, while gaps 2 and 3 are located at the high symmetry K and M points. Our edge state plot in Figure \ref{fig:surface&AHC}(b) shows high intensity in regions originally gapped by SOC in the bulk band structure, indicating electronic states and band crossings. Valence and conduction band crossings at these Weyl points are clearly observed in the edge state band structure (Figure \ref {fig:surface&AHC}(b)). The calculated topological invariant, \textit{C}, in the absence of external fields, is finite for the three bands, confirming the material as topologically nontrivial with three chiral modes terminating at the edge. Based on this analysis, we classify the ground state of 2D ScV$_6$Sn$_6$ as a Weyl semimetal with edge states near the Fermi energy, due to conduction-valence band crossings at discrete momentum points.~\cite{pokharel2021electronic,ortiz2021superconductivity}. Since TRS is inherently broken, the stability of the chiral edge states depends solely on the broken TRS and the lattice geometry. The weyl nodes appear in pairs of opposite chirality and are well separated in momentum space to prevent annihilation and ensure their stability. In the presence of SOC, 2D ScV$_6$Sn$_6$ exhibits a quantum anomalous Hall behavior, with the interplay between SOC and magnetic ordering offering additional stability to the chiral modes. For example, Fe$_3$Sn$_2$, a ferromagnetic kagome lattice, exhibits nontrivial topology due to strong SOC and magnetic order~\cite{yin2018giant}. Similarly, 2D ScV$_6$Sn$_6$ retains nontrivial topological characteristics protected by the kagome lattice structure and the strong interplay between SOC and intrinsic magnetic ordering. These features are comparable to those observed in the ferromagnetic Weyl semimetal Co$_3$Sn$_2$S$_2$~\cite{bernevig2022progress}.

\begin{figure*}[htb]
\centering
\includegraphics[scale = 0.8]{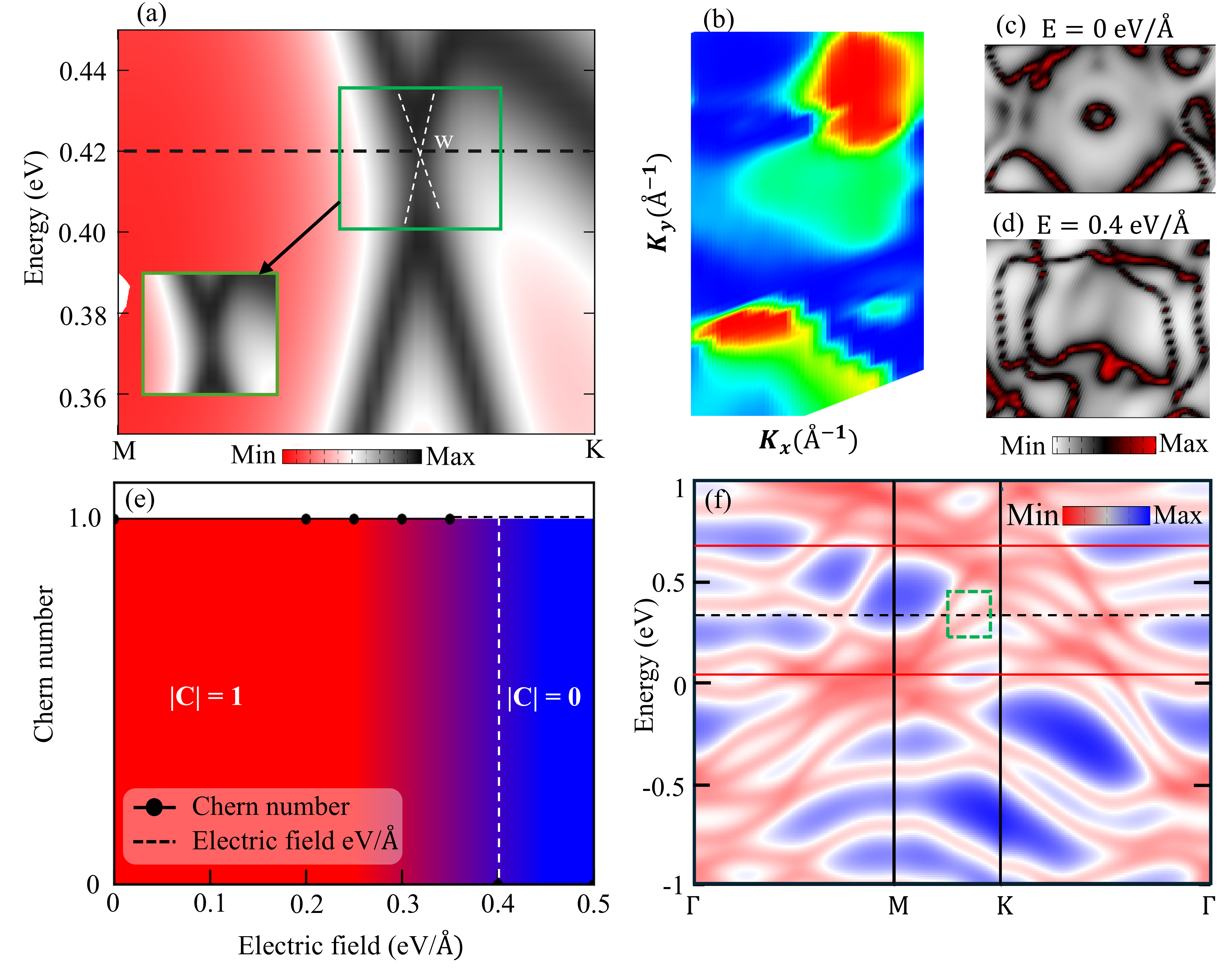}
\caption{(a) Topologically protected edge state in the TSM phase. The white dashed lines in the inset highlight the closing of the topological gap in the surface at `w`. (b) Berry curvature distributions of FM 2D ScV$_6$Sn$_6$ in the first Brillouin zone, with red regions marking Berry curvature singularities. (c)-(d) Fermi surface of the 110 slab: (c) without external fields and (d) with an electric field of 0.4 eV/\AA, highlighting changes in Fermi pockets at the critical \textit{E}-field. (e) The topological phase transition from the TSM phase ($|C|=1$) to the trivial phase ($|C|=0$), with the white dashed line showing the transition regime. (f) Edge states of FM 2D ScV$_6$Sn$_6$ (110) at a critical E-field of 0.4 eV/\AA, where red lines enclose bands containing nodal points. The dashed green box represents the position of the Weyl point $w$, while the dotted  black line indicates the Fermi level.}
 \label{fig:Fermi surfaces}
\end{figure*}

In ferromagnetic materials, the total anomalous Hall effect (AHE) conductivity, $\sigma_H^A$, arises from three main contributions: the intrinsic contribution $\sigma_{int}$, skew scattering $\sigma_{sc}$, and the side-jump mechanism $\sigma_{sj}$ \cite{chen2021large,nagaosa2010anomalous}. Mathematically, $\sigma_H^A$ is expressed as $\sigma_H^A = \sigma_{int} + \sigma_{sc} + \sigma_{sj}$. The intrinsic contribution, $\sigma_{int}$, depends on the material's band structure and is independent of scattering rates, while $\sigma_{sc}$ and $\sigma_{sj}$ originate from extrinsic scattering processes caused by impurities or disorder. We calculated the energy-dependent intrinsic AHC ($\sigma_{int}$) using the Berry curvature derived from band calculations. Figure \ref{fig:surface&AHC}(d) shows the AHC plot, with a peak near the Fermi energy, reaching 257 $\Omega^{-1}\text{cm}^{-1}$. The strong SOC and ferromagnetic ordering in 2D ScV$_6$Sn$_6$ enhance the Berry curvature, which acts as an effective magnetic field in momentum space, contributing to the large AHC. The peak at $\sim$0.33 eV near the Fermi energy can be attributed to the proximity of band crossings to the Fermi energy. The large anomalous Hall effect is believed to arise from the Berry curvature hotspots near the Fermi level. SOC-induced small gaps act as these Berry curvature hotspots, which in turn lead to the observed AHC. Currently, there are limited reports on AHC in 2D materials, making our findings particularly significant. This study provides valuable insights into the anomalous Hall effect in 2D kagome materials and contributes to the broader understanding of topological phenomena in these systems.

Previous studies on 2D ferromagnetic Dirac materials have reported intrinsic anomalous Hall conductivity (AHC) values reaching up to approximately 300 $\Omega^{-1}\text{cm}^{-1}$, driven by strong SOC and broken TRS \cite{yang2023tuning}. In Fe$_3$GeTe$_2$ monolayers, an intrinsic AHC of $\approx$350 $\Omega^{-1}\text{cm}^{-1}$ has been reported, with further enhancement achieved through external tuning methods such as ionic-liquid gating~\cite{guo2023modulating}. Similarly, studies on MnBi$_2$Te$_4$ have observed intrinsic AHC values ranging from 250 to 300 $\Omega^{-1}\text{cm}^{-1}$, attributed to contributions from Berry curvature and canted antiferromagnetic ordering. To the best of our knowledge, there are no previous reports on AHC in 2D kagome materials. Our predicted intrinsic AHC value of $\approx$257 $\Omega^{-1}\text{cm}^{-1}$ falls within the range of values reported for other 2D systems. Moreover, it is consistent with the values observed in bulk kagome materials, which range from $\sim$380 $\Omega^{-1}\text{cm}^{-1}$ for bulk LiMn$_6$Sn$_6$~\cite{chen2021large} to as high as 500 $\Omega^{-1} \text{cm}^{-1}$ for Fe$_3$Sn$_2$~\cite{belbase2023large}.

In Figure~\ref{fig:Fermi surfaces}, we investigate the impact of an externally applied perpendicular electric field on the topological properties of 2D ScV$_6$Sn$_6$. The edge states (Figures~\ref{fig:Fermi surfaces}(a)–(b)), associated with the Weyl point located near the Fermi energy, and the Berry curvature distribution at zero field are characterized by topological band crossings along the $M$–$K$ high-symmetry path in the Brillouin zone. Notably, the Berry curvature hotspots (Figure~\ref{fig:Fermi surfaces}(b)) coincide with Weyl point pairs both at and slightly below the Fermi level within the valence band. Upon applying an external electric field, the electronic structure undergoes systematic evolution. While low-intensity fields cause minimal perturbation, preserving the Weyl-like crossings and the overall topological character, higher field strengths induce significant band rearrangements. In particular, the chirality and location of Weyl points shift progressively, and a critical field strength of approximately 0.4~eV/\AA~marks the onset of a topological phase transition. At this threshold, spin–orbit coupling-driven bulk gaps emerge, and the previously gapless edge states begin to vanish, as shown in Figure~\ref{fig:Fermi surfaces}(f). The disappearance of topologically protected edge modes, along with a change in the Chern number from 1 to 0 (Figure~\ref{fig:Fermi surfaces}(e)), unambiguously confirms a transition from a nontrivial Weyl semimetal to a trivial metallic phase. Beyond this critical field, enhanced electronic coherence manifests through the formation of fully interconnected Fermi pockets (Figure~\ref{fig:Fermi surfaces}(d)), indicative of delocalized charge carriers in the trivial regime. Similar electric-field-induced topological transitions have been observed in other 2D materials, including MnBi$_2$Te$_4$, MoS$_2$, and WSe$_2$, where applied fields serve as effective tuning parameters to drive phase transitions between trivial and topological states~\cite{Qian_2014, Sawahata_2019, you2021electric}.  Although the system remains metallic across the entire range of applied electric field strengths, this field-induced evolution highlights a controllable pathway for tuning topological phases in 2D materials. Achieving a perpendicular electric field on the order of 0.4~eV/\AA~may be experimentally challenging via conventional gating methods; however, comparable field magnitudes can be realized through ionic-liquid gating in field-effect transistor architectures. Alternatively, substrate engineering employing polar or ferroelectric materials offers a promising route to modulate the local electrostatic potential and thereby manipulate the underlying band topology~\cite{weintrub2022generating, lv2019reconfigurable}.

In our simulations, we observed fluctuations in the Chern number at smaller electric fields, likely caused by modifications in the Berry curvature, driven by electric-field-induced alterations in the band structure. Although the precise mechanism behind these fluctuations remains unclear, similar behavior has been reported in MnBi$_2$Te$_4$, where the Chern number can be toggled by applying an electric field, with transitions between discrete values such as 1 and 2 at field strengths of around 1 to 2 eV/\AA{} \cite{Cai_2022}. Additionally, thin films near the topological critical point have exhibited fluctuations in the Chern number between 0, 1, and -1 as the electric field changes, eventually stabilizing at zero beyond approximately 2.5 eV/\AA{} \cite{lin2015quantum}. Rasouli et al. also observed similar electric-field-induced changes in the Chern number for 2D MnO$_2$, with stabilization at zero occurring beyond a certain threshold.

In contrast, the application of small magnetic fields (\textit{B}-fields) does not alter the topology of the bands. The Chern number remains unchanged regardless of the direction or magnitude of the applied magnetic field. This stability is due to the robustness of the topologically protected edge states, preserved by the system’s geometry and time-reversal symmetry (TRS). As moderate magnetic fields are insufficient to break this symmetry, the topological phase remains protected, similar to the role of spin-orbit coupling (SOC), which is also unable to break the symmetry \cite{moore2010birth}. Moreover, the weak coupling between the magnetic field and the electronic states in this material results in minimal effects on the band structure and topological properties. The stability of the Fermi surface under both in-plane and out-of-plane magnetic fields up to 2.1 T (not shown) further supports this robustness. The Fermi surface retains its 2D character, characterized by large hexagonal V-$d$ states and smaller circular Sc-$d$ states, as illustrated in Figures \ref{fig:Fermi surfaces}(b)-(c). Importantly, the shape of the Fermi surface remains consistent across different magnetic field strengths, indicating that the electronic structure is preserved under external magnetic fields. These findings are consistent with studies on other magnetic topological materials. For example, in MnBi$_2$Te$_4$, although magnetic fields induce transitions between antiferromagnetic (AFM) and ferromagnetic (FM) states, the Fermi surface remains largely unaffected, and no topological transition is observed \cite{tan2022mnbi2te4}. Similarly, in Cr-doped Bi$_2$Se$_3$ and 2D materials such as CrI$_3$ and Cr$_2$Ge$_2$Te$_2$, minimal topological changes occur under magnetic fields despite notable effects on magnetic properties \cite{nano13192655, liu2024magnetoresistance}.

\section{\label{sec:level4}Conclusions}

We have investigated the 2D kagome metal ScV$_6$Sn$_6$ and demonstrated that it exhibits topologically protected edge states and a Chern number of $|\textit{C}| = 1$ at the Fermi energy, confirming the presence of a Weyl semimetallic phase. This positions ScV$_6$Sn$_6$ among the few vanadium-based kagome compounds with nontrivial topology and broken time-reversal symmetry. Our analysis of edge states originating from gapped Weyl points near the Fermi energy reveals that the material retains metallic behavior at the surface despite being gapped in the bulk, confirming its topological nature. We further show that while magnetic fields do not affect its topology, an electric field of 0.4 eV/\AA{} induces a transition from a topological semimetal to a trivial metallic phase, with the edge states vanishing. This electrically induced transition, consistent with observations in other 2D materials, underscores the tunability of ScV$_6$Sn$_6$'s topological properties. Additionally, our calculations reveal a large anomalous Hall effect of $\sim$257 $\Omega^{-1}$cm$^{-1}$, driven by intrinsic Berry curvature and strong spin-orbit coupling, aligning with other 2D ferromagnetic Dirac-like materials. These findings highlight ScV$_6$Sn$_6$ as a pivotal material in 2D kagome systems, with strong potential for modulating topological properties in $d$-orbital-dominated structures. Its robust ferromagnetic ordering, tunable topological phases, and spin-orbit coupling make it an ideal candidate for advanced electronic and spintronic applications. The ability to control topological transitions through electric fields not only enhances our understanding of topological quantum materials but also paves the way for innovations in quantum computing and material design.

\section{\label{sec:level4}Acknowledgments}
This research is supported by the U.S. Department of Energy, Office of Science, Basic Energy Sciences under Award DOE-SC0024099. SKD was supported by the start up grant of Bitan Roy from Lehigh University. Computational resources were provided by Lehigh University Research Computing Infrastructure. We also acknowledge useful discussions with Benjamin Wieder.

\section{\label{sec:level4} Data availability}
Data will be made available upon reasonable request.

%\nocite{*}
%\bibliographystyle{plain}
%\bibliography{Ref}% Produces the bibliography via BibTeX.

\begin{thebibliography}{0}%
\makeatletter
\providecommand \@ifxundefined [1]{%
 \@ifx{#1\undefined}
}%
\providecommand \@ifnum [1]{%
 \ifnum #1\expandafter \@firstoftwo
 \else \expandafter \@secondoftwo
 \fi
}%
\providecommand \@ifx [1]{%
 \ifx #1\expandafter \@firstoftwo
 \else \expandafter \@secondoftwo
 \fi
}%
\providecommand \natexlab [1]{#1}%
\providecommand \enquote  [1]{``#1''}%
\providecommand \bibnamefont  [1]{#1}%
\providecommand \bibfnamefont [1]{#1}%
\providecommand \citenamefont [1]{#1}%
\providecommand \href@noop [0]{\@secondoftwo}%
\providecommand \href [0]{\begingroup \@sanitize@url \@href}%
\providecommand \@href[1]{\@@startlink{#1}\@@href}%
\providecommand \@@href[1]{\endgroup#1\@@endlink}%
\providecommand \@sanitize@url [0]{\catcode `\\12\catcode `\$12\catcode `\&12\catcode `\#12\catcode `\^12\catcode `\_12\catcode `\%12\relax}%
\providecommand \@@startlink[1]{}%
\providecommand \@@endlink[0]{}%
\providecommand \url  [0]{\begingroup\@sanitize@url \@url }%
\providecommand \@url [1]{\endgroup\@href {#1}{\urlprefix }}%
\providecommand \urlprefix  [0]{URL }%
\providecommand \Eprint [0]{\href }%
\providecommand \doibase [0]{http://dx.doi.org/}%
\providecommand \selectlanguage [0]{\@gobble}%
\providecommand \bibinfo  [0]{\@secondoftwo}%
\providecommand \bibfield  [0]{\@secondoftwo}%
\providecommand \translation [1]{[#1]}%
\providecommand \BibitemOpen [0]{}%
\providecommand \bibitemStop [0]{}%
\providecommand \bibitemNoStop [0]{.\EOS\space}%
\providecommand \EOS [0]{\spacefactor3000\relax}%
\providecommand \BibitemShut  [1]{\csname bibitem#1\endcsname}%
\let\auto@bib@innerbib\@empty
%</preamble>
\end{thebibliography}%


\begin{thebibliography}{10}

\bibitem{hasan2015topological}
M~Zahid Hasan, Su-Yang Xu, and Guang Bian.
\newblock Topological insulators, topological superconductors and weyl fermion semimetals: discoveries, perspectives and outlooks.
\newblock {\em Physica Scripta}, 2015(T164):014001, 2015.

\bibitem{RevModPhys.82.3045}
M.~Z. Hasan and C.~L. Kane.
\newblock Colloquium: Topological insulators.
\newblock {\em Rev. Mod. Phys.}, 82:3045--3067, Nov 2010.

\bibitem{kane2005quantum}
Charles~L Kane and Eugene~J Mele.
\newblock Quantum spin hall effect in graphene.
\newblock {\em Physical review letters}, 95(22):226801, 2005.

\bibitem{niu2015two}
Chengwang Niu, Patrick~M Buhl, Gustav Bihlmayer, Daniel Wortmann, Stefan Blugel, and Yuriy Mokrousov.
\newblock Two-dimensional topological crystalline insulator and topological phase transition in tlse and tls monolayers.
\newblock {\em Nano letters}, 15(9):6071--6075, 2015.

\bibitem{wang2013organic}
ZF~Wang, Zheng Liu, and Feng Liu.
\newblock Organic topological insulators in organometallic lattices.
\newblock {\em Nature communications}, 4(1):1471, 2013.

\bibitem{okada2013observation}
Yoshinori Okada, Maksym Serbyn, Hsin Lin, Daniel Walkup, Wenwen Zhou, Chetan Dhital, Madhab Neupane, Suyang Xu, Yung~Jui Wang, Raman Sankar, et~al.
\newblock Observation of dirac node formation and mass acquisition in a topological crystalline insulator.
\newblock {\em Science}, 341(6153):1496--1499, 2013.

\bibitem{xu2020high}
Yuanfeng Xu, Luis Elcoro, Zhi-Da Song, Benjamin~J Wieder, MG~Vergniory, Nicolas Regnault, Yulin Chen, Claudia Felser, and B~Andrei Bernevig.
\newblock High-throughput calculations of magnetic topological materials.
\newblock {\em Nature}, 586(7831):702--707, 2020.

\bibitem{watanabe2018structure}
Haruki Watanabe, Hoi~Chun Po, and Ashvin Vishwanath.
\newblock Structure and topology of band structures in the 1651 magnetic space groups.
\newblock {\em Science advances}, 4(8):eaat8685, 2018.

\bibitem{otrokov2019prediction}
Mikhail~M Otrokov, Ilya~I Klimovskikh, Hendrik Bentmann, D~Estyunin, Alexander Zeugner, Ziya~S Aliev, S~Ga{\ss}, AUB Wolter, AV~Koroleva, Alexander~M Shikin, et~al.
\newblock Prediction and observation of an antiferromagnetic topological insulator.
\newblock {\em Nature}, 576(7787):416--422, 2019.

\bibitem{sheng2006quantum}
DN~Sheng, ZY~Weng, L~Sheng, and FDM Haldane.
\newblock Quantum spin-hall effect and topologically invariant chern numbers.
\newblock {\em Physical review letters}, 97(3):036808, 2006.

\bibitem{prodan2010entanglement}
Emil Prodan, Taylor~L Hughes, and B~Andrei Bernevig.
\newblock Entanglement spectrum of a disordered topological chern insulator.
\newblock {\em Physical review letters}, 105(11):115501, 2010.

\bibitem{ahsan2023prediction}
Taosif Ahsan, Chia-Hsiu Hsu, Md~Shafayat Hossain, and M~Zahid Hasan.
\newblock Prediction of strong topological insulator phase in kagome metal r v 6 ge 6.
\newblock {\em Physical Review Materials}, 7(10):104204, 2023.

\bibitem{teng2022discovery}
Xiaokun Teng, Lebing Chen, Feng Ye, Elliott Rosenberg, Zhaoyu Liu, Jia-Xin Yin, Yu-Xiao Jiang, Ji~Seop Oh, M~Zahid Hasan, Kelly~J Neubauer, et~al.
\newblock Discovery of charge density wave in a kagome lattice antiferromagnet.
\newblock {\em Nature}, 609(7927):490--495, 2022.

\bibitem{tan2021charge}
Hengxin Tan, Yizhou Liu, Ziqiang Wang, and Binghai Yan.
\newblock Charge density waves and electronic properties of superconducting kagome metals.
\newblock {\em Physical review letters}, 127(4):046401, 2021.

\bibitem{yu2021concurrence}
FH~Yu, T~Wu, ZY~Wang, B~Lei, WZ~Zhuo, JJ~Ying, and XH~Chen.
\newblock Concurrence of anomalous hall effect and charge density wave in a superconducting topological kagome metal.
\newblock {\em Physical Review B}, 104(4):L041103, 2021.

\bibitem{hu2022tunable}
Yong Hu, Xianxin Wu, Yongqi Yang, Shunye Gao, Nicholas~C Plumb, Andreas~P Schnyder, Weiwei Xie, Junzhang Ma, and Ming Shi.
\newblock Tunable topological dirac surface states and van hove singularities in kagome metal gdv6sn6.
\newblock {\em Science Advances}, 8(38):eadd2024, 2022.

\bibitem{pokharel2021electronic}
Ganesh Pokharel, Samuel~ML Teicher, Brenden~R Ortiz, Paul~M Sarte, Guang Wu, Shuting Peng, Junfeng He, Ram Seshadri, and Stephen~D Wilson.
\newblock Electronic properties of the topological kagome metals yv 6 sn 6 and gdv 6 sn 6.
\newblock {\em Physical Review B}, 104(23):235139, 2021.

\bibitem{morali2019fermi}
Noam Morali, Rajib Batabyal, Pranab~Kumar Nag, Enke Liu, Qiunan Xu, Yan Sun, Binghai Yan, Claudia Felser, Nurit Avraham, and Haim Beidenkopf.
\newblock Fermi-arc diversity on surface terminations of the magnetic weyl semimetal co3sn2s2.
\newblock {\em Science}, 365(6459):1286--1291, 2019.

\bibitem{hirschberger2016chiral}
Max Hirschberger, Satya Kushwaha, Zhijun Wang, Quinn Gibson, Sihang Liang, Carina~A Belvin, Bogdan~Andrei Bernevig, Robert~J Cava, and Nai~Phuan Ong.
\newblock The chiral anomaly and thermopower of weyl fermions in the half-heusler gdptbi.
\newblock {\em Nature materials}, 15(11):1161--1165, 2016.

\bibitem{kubler2014non}
J{\"u}rgen K{\"u}bler and Claudia Felser.
\newblock Non-collinear antiferromagnets and the anomalous hall effect.
\newblock {\em Europhysics Letters}, 108(6):67001, 2014.

\bibitem{zhang2017strong}
Yang Zhang, Yan Sun, Hao Yang, Jakub {\v{Z}}elezn{\`y}, Stuart~PP Parkin, Claudia Felser, and Binghai Yan.
\newblock Strong anisotropic anomalous hall effect and spin hall effect in the chiral antiferromagnetic compounds mn 3 x (x= ge, sn, ga, ir, rh, and pt).
\newblock {\em Physical Review B}, 95(7):075128, 2017.

\bibitem{aliev2019novel}
Ziya~S Aliev, Imamaddin~R Amiraslanov, Daria~I Nasonova, Andrei~V Shevelkov, Nadir~A Abdullayev, Zakir~A Jahangirli, Elnur~N Orujlu, Mikhail~M Otrokov, Nazim~T Mamedov, Mahammad~B Babanly, et~al.
\newblock Novel ternary layered manganese bismuth tellurides of the mnte-bi2te3 system: Synthesis and crystal structure.
\newblock {\em Journal of Alloys and Compounds}, 789:443--450, 2019.

\bibitem{Berry1984quantal}
Michael~Victor Berry.
\newblock Quantal phase factors accompanying adiabatic changes.
\newblock {\em Proceedings of the Royal Society of London. A. Mathematical and Physical Sciences}, 392(1802):45--57, 1984.

\bibitem{Sawahata_2019}
Hikaru Sawahata, Naoya Yamaguchi, and Fumiyuki Ishii.
\newblock Electric-field-induced z2 topological phase transition in strained single bilayer bi(111).
\newblock {\em Applied Physics Express}, 12(7):075009, jun 2019.

\bibitem{liu2015switching}
Qihang Liu, Xiuwen Zhang, LB~Abdalla, Adalberto Fazzio, and Alex Zunger.
\newblock Switching a normal insulator into a topological insulator via electric field with application to phosphorene.
\newblock {\em Nano letters}, 15(2):1222--1228, 2015.

\bibitem{sun2022valley}
Hao Sun, Sheng-Shi Li, Wei-xiao Ji, and Chang-Wen Zhang.
\newblock Valley-dependent topological phase transition and quantum anomalous valley hall effect in single-layer ruclbr.
\newblock {\em Physical Review B}, 105(19):195112, 2022.

\bibitem{wu2023quantum}
Bin Wu, Yong-liang Song, Wei-xiao Ji, Pei-ji Wang, Shu-feng Zhang, and Chang-wen Zhang.
\newblock Quantum anomalous hall effect in an antiferromagnetic monolayer of moo.
\newblock {\em Physical Review B}, 107(21):214419, 2023.

\bibitem{arachchige2022charge}
Hasitha W~Suriya Arachchige, William~R Meier, Madalynn Marshall, Takahiro Matsuoka, Rui Xue, Michael~A McGuire, Raphael~P Hermann, Huibo Cao, and David Mandrus.
\newblock Charge density wave in kagome lattice intermetallic scv 6 sn 6.
\newblock {\em Physical Review Letters}, 129(21):216402, 2022.

\bibitem{destefano2023pseudogap}
Jonathan~M DeStefano, Elliott Rosenberg, Olivia Peek, Yongbin Lee, Zhaoyu Liu, Qianni Jiang, Liqin Ke, and Jiun-Haw Chu.
\newblock Pseudogap behavior in charge density wave kagome material scv6sn6 revealed by magnetotransport measurements.
\newblock {\em npj Quantum Materials}, 8(1):65, 2023.

\bibitem{chang2023colloquium}
Cui-Zu Chang, Chao-Xing Liu, and Allan~H MacDonald.
\newblock Colloquium: Quantum anomalous hall effect.
\newblock {\em Reviews of Modern Physics}, 95(1):011002, 2023.

\bibitem{vzelezny2023high}
Jakub {\v{Z}}elezn{\`y}, Yuta Yahagi, Carles Gomez-Olivella, Yang Zhang, and Yan Sun.
\newblock High-throughput study of the anomalous hall effect.
\newblock {\em npj Computational Materials}, 9(1):151, 2023.

\bibitem{ye2018massive}
Linda Ye, Mingu Kang, Junwei Liu, Felix Von~Cube, Christina~R Wicker, Takehito Suzuki, Chris Jozwiak, Aaron Bostwick, Eli Rotenberg, David~C Bell, et~al.
\newblock Massive dirac fermions in a ferromagnetic kagome metal.
\newblock {\em Nature}, 555(7698):638--642, 2018.

\bibitem{kohn1996density}
Walter Kohn, Axel~D Becke, and Robert~G Parr.
\newblock Density functional theory of electronic structure.
\newblock {\em The journal of physical chemistry}, 100(31):12974--12980, 1996.

\bibitem{kresse1996efficiency}
Georg Kresse and J{\"u}rgen Furthm{\"u}ller.
\newblock Efficiency of ab-initio total energy calculations for metals and semiconductors using a plane-wave basis set.
\newblock {\em Computational materials science}, 6(1):15--50, 1996.

\bibitem{kresse1996efficient}
Georg Kresse and J{\"u}rgen Furthm{\"u}ller.
\newblock Efficient iterative schemes for ab initio total-energy calculations using a plane-wave basis set.
\newblock {\em Physical review B}, 54(16):11169, 1996.

\bibitem{perdew1996generalized}
John~P Perdew, Kieron Burke, and Matthias Ernzerhof.
\newblock Generalized gradient approximation made simple.
\newblock {\em Physical review letters}, 77(18):3865, 1996.

\bibitem{anisimov2005full}
VI~Anisimov, DE~Kondakov, AV~Kozhevnikov, IA~Nekrasov, ZV~Pchelkina, JW~Allen, S-K Mo, H-D Kim, P~Metcalf, S~Suga, et~al.
\newblock Full orbital calculation scheme for materials with strongly correlated electrons.
\newblock {\em Physical Review B}, 71(12):125119, 2005.

\bibitem{BERDECIA2024113067}
B.H. Berdecia and C.E. Ekuma.
\newblock Electronic and optical properties of yb-based 1-2-20 materials.
\newblock {\em Computational Materials Science}, 242:113067, 2024.

\bibitem{Marzari_2012}
Nicola Marzari, Arash~A. Mostofi, Jonathan~R. Yates, Ivo Souza, and David Vanderbilt.
\newblock Maximally localized wannier functions: Theory and applications.
\newblock {\em Reviews of Modern Physics}, 84(4):1419–1475, October 2012.

\bibitem{Pizzi2020}
Giovanni Pizzi, Valerio Vitale, Ryotaro Arita, Stefan Blügel, Frank Freimuth, Guillaume G{\'{e}}ranton, Marco Gibertini, Dominik Gresch, Charles Johnson, Takashi Koretsune, Julen Iba{\~{n}}ez-Azpiroz, Hyungjun Lee, Jae-Mo Lihm, Daniel Marchand, Antimo Marrazzo, Yuriy Mokrousov, Jamal~I Mustafa, Yoshiro Nohara, Yusuke Nomura, Lorenzo Paulatto, Samuel Ponc{\'{e}}, Thomas Ponweiser, Junfeng Qiao, Florian Thöle, Stepan~S Tsirkin, Ma{\l}gorzata Wierzbowska, Nicola Marzari, David Vanderbilt, Ivo Souza, Arash~A Mostofi, and Jonathan~R Yates.
\newblock Wannier90 as a community code: new features and applications.
\newblock {\em Journal of Physics: Condensed Matter}, 32(16):165902, jan 2020.

\bibitem{WU2017}
QuanSheng Wu, ShengNan Zhang, Hai-Feng Song, Matthias Troyer, and Alexey~A. Soluyanov.
\newblock Wanniertools : An open-source software package for novel topological materials.
\newblock {\em Computer Physics Communications}, 224:405 -- 416, 2018.

\bibitem{wang2015measurement}
Wen Wang, Shuyang Dai, Xide Li, Jiarui Yang, David~J Srolovitz, and Quanshui Zheng.
\newblock Measurement of the cleavage energy of graphite.
\newblock {\em Nature communications}, 6(1):1--7, 2015.

\bibitem{coleman2011two}
Jonathan~N Coleman, Mustafa Lotya, Arlene O’Neill, Shane~D Bergin, Paul~J King, Umar Khan, Karen Young, Alexandre Gaucher, Sukanta De, Ronan~J Smith, et~al.
\newblock Two-dimensional nanosheets produced by liquid exfoliation of layered materials.
\newblock {\em Science}, 331(6017):568--571, 2011.

\bibitem{tang2014nanomechanical}
Dai-Ming Tang, Dmitry~G Kvashnin, Sina Najmaei, Yoshio Bando, Koji Kimoto, Pekka Koskinen, Pulickel~M Ajayan, Boris~I Yakobson, Pavel~B Sorokin, Jun Lou, et~al.
\newblock Nanomechanical cleavage of molybdenum disulphide atomic layers.
\newblock {\em Nature communications}, 5(1):3631, 2014.

\bibitem{gaur2014surface}
Anand~PS Gaur, Satyaprakash Sahoo, Majid Ahmadi, Saroj~P Dash, Maxime J-F Guinel, and Ram~S Katiyar.
\newblock Surface energy engineering for tunable wettability through controlled synthesis of mos2.
\newblock {\em Nano letters}, 14(8):4314--4321, 2014.

\bibitem{liu2022elastool}
Zhong-Li Liu, CE~Ekuma, Wei-Qi Li, Jian-Qun Yang, and Xing-Ji Li.
\newblock Elastool: An automated toolkit for elastic constants calculation.
\newblock {\em Computer Physics Communications}, 270:108180, 2022.

\bibitem{EKUMA2024109161}
C.E. Ekuma and Z.-L. Liu.
\newblock Elastool v3.0: Efficient computational and visualization toolkit for elastic and mechanical properties of materials.
\newblock {\em Computer Physics Communications}, 300:109161, 2024.

\bibitem{born1996dynamical}
Max Born and Kun Huang.
\newblock {\em Dynamical theory of crystal lattices}.
\newblock Oxford university press, 1996.

\bibitem{PhysRevB.90.224104}
F\'elix Mouhat and Fran\ifmmode \mbox{\c{c}}\else \c{c}\fi{}ois-Xavier Coudert.
\newblock Necessary and sufficient elastic stability conditions in various crystal systems.
\newblock {\em Phys. Rev. B}, 90:224104, Dec 2014.

\bibitem{supp}
See Supplemental Material at URL-will-be-inserted-by-publisher.

\bibitem{di2023electronic}
Domenico Di~Sante, Bongjae Kim, Werner Hanke, Tim Wehling, Cesare Franchini, Ronny Thomale, and Giorgio Sangiovanni.
\newblock Electronic correlations and universal long-range scaling in kagome metals.
\newblock {\em Physical Review Research}, 5(1):L012008, 2023.

\bibitem{yi2024quantum}
Changjiang Yi, Xiaolong Feng, Ning Mao, Premakumar Yanda, Subhajit Roychowdhury, Yang Zhang, Claudia Felser, and Chandra Shekhar.
\newblock Quantum oscillations revealing topological band in kagome metal scv 6 sn 6.
\newblock {\em Physical Review B}, 109(3):035124, 2024.

\bibitem{yi2024tuning}
Changjiang Yi, Xiaolong Feng, Nitesh Kumar, Claudia Felser, and Chandra Shekhar.
\newblock Tuning charge density wave of kagome metal scv6sn6, 2024.

\bibitem{ortiz2021superconductivity}
Brenden~R Ortiz, Paul~M Sarte, Eric~M Kenney, Michael~J Graf, Samuel~ML Teicher, Ram Seshadri, and Stephen~D Wilson.
\newblock Superconductivity in the z 2 kagome metal kv 3 sb 5.
\newblock {\em Physical Review Materials}, 5(3):034801, 2021.

\bibitem{yin2018giant}
Jia-Xin Yin, Songtian~S Zhang, Hang Li, Kun Jiang, Guoqing Chang, Bingjing Zhang, Biao Lian, Cheng Xiang, Ilya Belopolski, Hao Zheng, et~al.
\newblock Giant and anisotropic many-body spin--orbit tunability in a strongly correlated kagome magnet.
\newblock {\em Nature}, 562(7725):91--95, 2018.

\bibitem{bernevig2022progress}
B~Andrei Bernevig, Claudia Felser, and Haim Beidenkopf.
\newblock Progress and prospects in magnetic topological materials.
\newblock {\em Nature}, 603(7899):41--51, 2022.

\bibitem{chen2021large}
Dong Chen, Congcong Le, Chenguang Fu, Haicheng Lin, Walter Schnelle, Yan Sun, and Claudia Felser.
\newblock Large anomalous hall effect in the kagome ferromagnet limn 6 sn 6.
\newblock {\em Physical Review B}, 103(14):144410, 2021.

\bibitem{nagaosa2010anomalous}
Naoto Nagaosa, Jairo Sinova, Shigeki Onoda, Allan~H MacDonald, and Nai~Phuan Ong.
\newblock Anomalous hall effect.
\newblock {\em Reviews of modern physics}, 82(2):1539, 2010.

\bibitem{yang2023tuning}
Xin Yang, Yanqing Shen, Lingling Lv, Min Zhou, Yu~Zhang, Xianghui Meng, Xiangqian Jiang, Qing Ai, Yong Shuai, and Zhongxiang Zhou.
\newblock Tuning the topological phase and anomalous hall conductivity with magnetization direction in h-fecl2 monolayer.
\newblock {\em Applied Physics Letters}, 123(20), 2023.

\bibitem{guo2023modulating}
Min Guo, Ju~Zhou, Hai-Shuang Lu, Sheng Ju, and Tian-Yi Cai.
\newblock Modulating intrinsic anomalous hall effect in fe3gete2 monolayer via strain engineering.
\newblock {\em AIP Advances}, 13(10), 2023.

\bibitem{belbase2023large}
Bishnu~P Belbase, Linda Ye, Bishnu Karki, Jorge~I Facio, Jhih-Shih You, Joseph~G Checkelsky, Jeroen Van Den~Brink, and Madhav~Prasad Ghimire.
\newblock Large anomalous hall effect in single crystals of the kagome weyl ferromagnet fe 3 sn.
\newblock {\em Physical Review B}, 108(7):075164, 2023.

\bibitem{Qian_2014}
Xiaofeng Qian, Junwei Liu, Liang Fu, and Ju~Li.
\newblock Quantum spin hall effect in two-dimensional transition metal dichalcogenides.
\newblock {\em Science}, 346(6215):1344–1347, December 2014.

\bibitem{you2021electric}
Jing-Yang You, Xue-Juan Dong, Bo~Gu, and Gang Su.
\newblock Electric field induced topological phase transition and large enhancements of spin-orbit coupling and curie temperature in two-dimensional ferromagnetic semiconductors.
\newblock {\em Physical Review B}, 103(10):104403, 2021.

\bibitem{weintrub2022generating}
Benjamin~I Weintrub, Yu-Ling Hsieh, Sviatoslav Kovalchuk, Jan~N Kirchhof, Kyrylo Greben, and Kirill~I Bolotin.
\newblock Generating intense electric fields in 2d materials by dual ionic gating.
\newblock {\em Nature Communications}, 13(1):6601, 2022.

\bibitem{lv2019reconfigurable}
Liang Lv, Fuwei Zhuge, Fengjun Xie, Xujing Xiong, Qingfu Zhang, Nan Zhang, Yu~Huang, and Tianyou Zhai.
\newblock Reconfigurable two-dimensional optoelectronic devices enabled by local ferroelectric polarization.
\newblock {\em Nature communications}, 10(1):3331, 2019.

\bibitem{Cai_2022}
Jiaqi Cai, Dmitry Ovchinnikov, Zaiyao Fei, Minhao He, Tiancheng Song, Zhong Lin, Chong Wang, David Cobden, Jiun-Haw Chu, Yong-Tao Cui, Cui-Zu Chang, Di~Xiao, Jiaqiang Yan, and Xiaodong Xu.
\newblock Electric control of a canted-antiferromagnetic chern insulator.
\newblock {\em Nature Communications}, 13(1), March 2022.

\bibitem{lin2015quantum}
Hsin Lin, Wei-Feng Tsai, YP~Feng, et~al.
\newblock Quantum anomalous hall effect with field-tunable chern number near z 2 topological critical point.
\newblock {\em Physical Review B}, 92(11):115205, 2015.

\bibitem{moore2010birth}
Joel~E Moore.
\newblock The birth of topological insulators.
\newblock {\em Nature}, 464(7286):194--198, 2010.

\bibitem{tan2022mnbi2te4}
Weilun Tan, Jing Liu, Hui Li, Dandan Guan, and Jin-Feng Jia.
\newblock Mnbi2te4--a good platform for topological quantum physics study.
\newblock {\em Quantum Frontiers}, 1(1):19, 2022.

\bibitem{nano13192655}
Gang Qiu, Hung-Yu Yang, Su~Kong Chong, Yang Cheng, Lixuan Tai, and Kang~L. Wang.
\newblock Manipulating topological phases in magnetic topological insulators.
\newblock {\em Nanomaterials}, 13(19), 2023.

\bibitem{liu2024magnetoresistance}
Han-lei Liu, Zi-yan Luo, Jun-jie Guo, Xi-guang Wang, Yao-zhuang Nie, Qing-lin Xia, and Guang-hua Guo.
\newblock Magnetoresistance and magnetic field-induced phase transition in two-dimensional antiferromagnet fe1/3nbs2.
\newblock {\em AIP Advances}, 14(5), 2024.

\end{thebibliography}

%\bibliographystyle{unsrt}

\end{document}

\LARGE
\textbf{Supplementary Materials: Tunable topological phase in 2D ScV$_6$Sn$_6$ kagome material}\footnote{This research is supported by the U.S. Department of Energy, Office of Science, Basic Energy Sciences under Award DOE-SC0024099 (algorithmic development and first-principles calculations) and the Lehigh Core grant (initial high-throughput framework development). SKD was supported by the start-up grant of Bitan Roy from Lehigh University. Computational resources were provided by Lehigh University Research Computing Infrastructure. We also acknowledge useful discussions with Benjamin Wieder.}\\[6pt]
\small
\textbf {Chidiebere I. Nwaogbo${^1}$, Sanjib K. Das${^{1,2}}$ and  Chinedu E. Ekuma${^1}$}\\[6pt]
Department of Physics, Lehigh University, Bethlehem, PA USA${^1}$\\
Department of Physics and Astronomy, University of Delaware, Newark, DE 19716, USA${^2}$\\ cin221@lehigh.edu${^1}$\\skdas@udel.edu${^2}$\\ che218@lehigh.edu${^1}$\\[6pt]

%a.\textit{\textbf{ Structural stability}}

\begin{figure}[htb!]
	\centering
	\includegraphics[trim = 0mm 0mm 0mm 0mm,width=1.0\linewidth,clip=true]{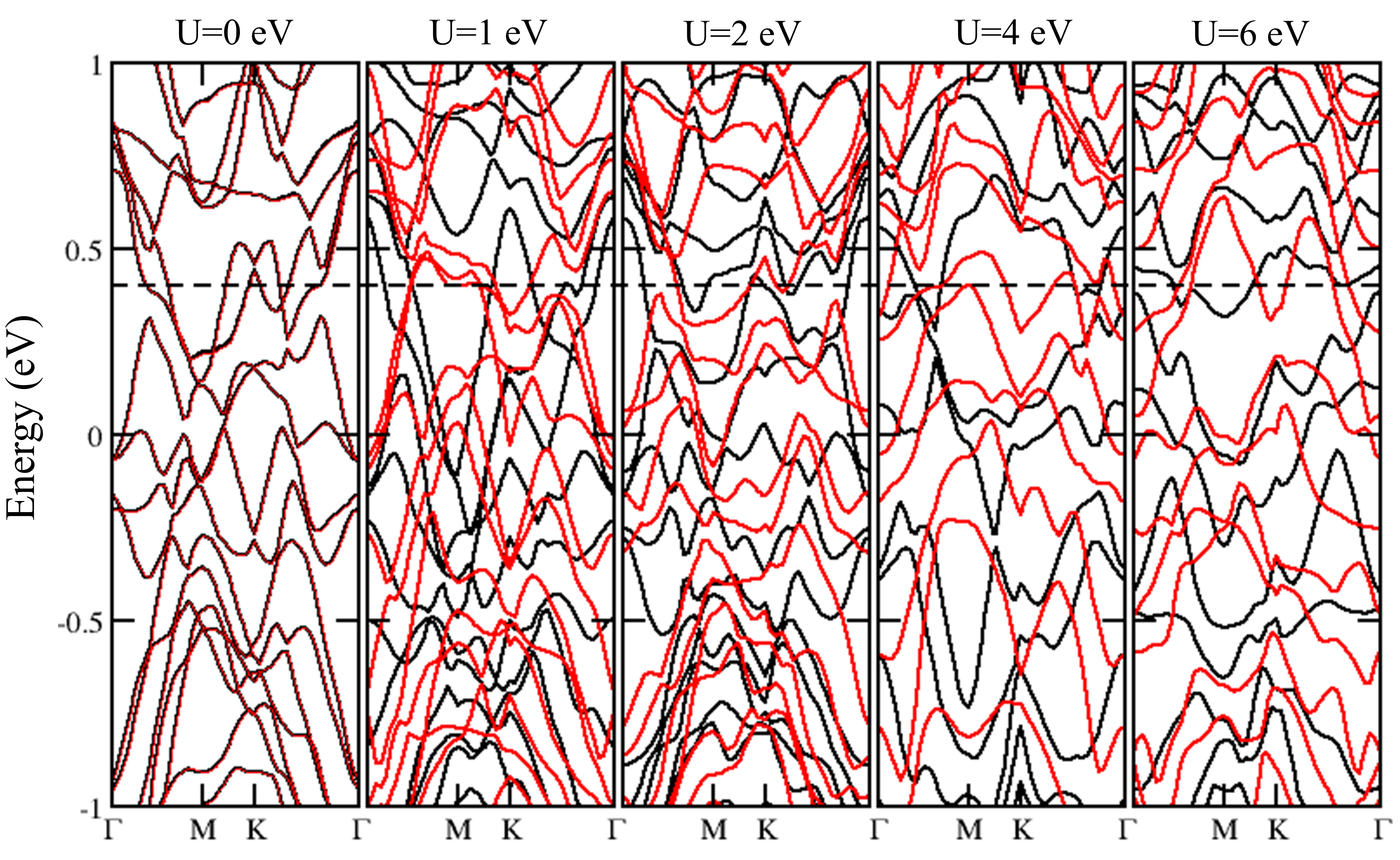} 
	\caption{Spin-polarized electronic band structures of 2D ScV$_6$Sn$_6$ computed for different values of the Hubbard \( U \) parameter applied to the V 3\textit{d} and Sc orbitals. The results demonstrate that the overall band topology, including the metallic character and Weyl-like crossings near the Fermi level, remains qualitatively unchanged across the full range of \( U \) values, confirming the robustness of the electronic structure against on-site Coulomb interactions.} 
	\label{figS1}
\end{figure}

To evaluate the sensitivity of the electronic and topological properties of 2D ScV$_6$Sn$_6$ to the choice of the Hubbard \textit{U}  parameter, we performed spin-polarized DFT+U calculations by systematically varying the on-site Coulomb interaction applied to the V 3d orbitals from 0 to 6 eV. For consistency, the same  U value was applied to the Sc 3d states, although their contribution near the Fermi level is relatively minor. Figure~\ref{figS1}  shows the corresponding band structures for each  \textit{U} value. At \textit{U} = 0 eV, the system exhibits a nearly degenerate spin state with very weak spin splitting, indicating marginal ferromagnetic behavior in the absence of on-site Coulomb interaction. As U increases, spin splitting becomes more pronounced, and the magnetic moment on the V atoms stabilizes, confirming the robustness of the ferromagnetic ground state. Importantly, the key topological features such as spin-orbit-induced band gaps, Weyl-like crossings near the Fermi energy, and the presence of chiral edge modes remain qualitatively unchanged across the entire range of \textit{U} values. This consistency demonstrates that the topological phase and associated magnetic ordering are intrinsic properties of the system and are not artifacts of a specific choice of \textit{U}. Therefore, our originally chosen value of \textit{U} = 6 eV, guided by prior studies on vanadium-based kagome systems, is well justified and captures the essential physics of the 2D  ScV$_6$Sn$_6$ kagome material.

Regarding the K-point sampling for all calculations, we performed using a $5\times5\times1$ gamma-centered grid, consistent with established practices for large 2D unit cells. Convergence tests with denser $7\times7\times1$ and $9\times9\times1$ grids showed total energy differences of less than 0.01 eV/atom and negligible changes in the band structure, confirming the adequacy of the $5\times5\times1$ grid for accurately capturing the electronic and topological properties of 2D  ScV$_6$Sn$_6$.

%\nocite{*}
%\bibliographystyle{plain}
%\bibliography{Ref}% Produces the bibliography via BibTeX.